%% file: manuscript_R1.tex
\documentclass[12pt]{iopart}

\pdfoutput=1

\usepackage{graphicx}
\usepackage{harvard}

\input{format.ltx}

\newcommand{\MT}{M-T}
\newcommand{\FEM}{FEM}
\newcommand{\EPUC}{EPUC}
\newcommand{\PMM}{PMM}
\newcommand{\CC}{C/C}

\newcommand{\figName}[1]{\includegraphics{#1}}
\newcommand{\code}[1]{{\sc #1}}

\newcommand{\sa}{\xi}
\newcommand{\oavg}[1]{\left\langle\!\left\langle #1 \right\rangle\!\right\rangle}

\usepackage{color}

\definecolor{myGray}{rgb}{.75,.75,.75}

\newcommand{\new}[1]{#1}
\newcommand{\delete}[1]{}
\newcommand{\replace}[2]{\delete{#1}\new{#2}}

\begin{document}

\title[Effective Properties of \delete{Carbon-Carbon} Textile
  Composites]{Effective Properties of \delete{Carbon-Carbon} Textile
  Composites: Application of the Mori-Tanaka Method}
\author{Jan Sko\v{c}ek$^1$, Jan Zeman$^2$, Michal \v{S}ejnoha$^{2,3}$}
\address{$^1$ Department of Civil Engineering, Technical University of
  Denmark, Brovej, Building 118, DK-2800 Kgs. Lyngby, Denmark}
\ead{jas@byg.dtu.dk}
\address{$^2$ Department of Mechanics, Faculty of Civil
  Engineering, Czech Technical University in Prague, Th\' akurova 7,
  166 29 Prague 6, Czech Republic}
\ead{zemanj@cml.fsv.cvut.cz}
\address{$^3$ Centre for Integrated Design of Advances Structures,
  Th\' akurova 7, 166 29 Prague 6, Czech Republic}
\ead{sejnom@fsv.cvut.cz}

\begin{abstract}
An efficient approach to the evaluation of effective elastic
properties of \delete{carbon-carbon} plain weave textile composites
using the Mori-Tanaka method is presented. The method proves its
potential even if applied to real material systems with various types
of imperfections including the non-uniform waviness of the fiber-tow
paths, both along its longitudinal direction and through the laminate
thickness. Influence of the remaining geometrical parameters is
accounted for by optimal calibration of the shape of the equivalent
ellipsoidal inclusion. An application of the method to a particular
sample of the carbon-carbon composite laminate demonstrates not only
its applicability but also its efficiency particularly when compared
to finite element simulations.
\end{abstract}


\submitto{\MSMSE}

\maketitle

\section{Introduction}\label{sec:intro}
%%%%%%%%%%%%%%%%%%%%%%%%%%%%%%%%%%%%%%%%%%%%%%%%%%%%%%%%%%%%%%%%%%%%%%%%%%%%%%%%%%%%%%%%%%%%%%%%%%%%%%%%%%%%%%%%%%%%%%%%%%%%%
\replace{Carbon-carbon \CC~p}{P}lain weave composites, reinforced by
mutually interlaced systems of unidirectional \delete{carbon} fiber
tows bonded to a \delete{carbon} matrix, belong to a progressive
material systems \replace{many applications not only in the space and
  automobile industry, but also in the medicine owing to its appealing
  biological compatibility with a living soft tissue}{with widespread
  applications in virtually all areas of engineering. As a particular
  example, consider carbon-carbon (\CC)~composites, originally developed
  for the space and automobile industry, which now find uses in the
  medicine owing to their appealing biological compatibility with a
  living soft tissue, e.g.~\cite{Pesakova:2003:BBP}}. A proper characterization of
these material systems, especially from the mechanical response point
of view, thus appears rather important.

While a detailed two-dimensional (2D) analysis of a heat conduction
problem for the evaluation of effective (macroscopic) thermal
conductivities seems to be sufficient,
\new{e.g.}~\cite{Tomkova:2006,Tomkova:2008:EETC}, a reliable estimate
of the mechanical response of such systems requires in general a
solution of a full three-dimensional (3D) problem.  This task,
however, presents a significant challenge even if limiting our
attention to a linear elastic behavior. Not only the characteristic 3D
structure of textile composites, but also various types of
imperfections in woven path developed during the manufacturing process
preclude a direct formulation of a simple computational model.

A considerable research effort has been invested in the last two
decades into providing a simple yet accurate scheme for the
predictions of macroscopic elastic properties of woven
composites. With an increasing level of sophistication, these models
include modified rule of mixtures, approaches based on classical
laminate theories~(CLT) and detailed three-dimensional finite element
method~(FEM) based simulations, see
e.g.~\cite{Cox:1997:HAMTC,Chung:1999:WFC,Takano:1999:HMTC,Lomov:2007:MFEM}
for a review and comparison of individual approaches. The latter class
of computational models is considered to be the most accurate one
particularly if combined with concise geometrical
data~\cite{Barbero:2006:FEM,Zeman:2004:RC,Lomov:2007:MFEM}.

The FEM simulations show, however, certain disadvantages.  Perhaps the
most critical one is a relatively high computational cost due to
laborious preparation of finite element meshes. Moreover, to
incorporate at least the dominant microstructural imperfections
observed in real systems into the FEM model is far from being trivial
and deserves a special treatment typically based on an appropriate
statistical
characterization~\cite{Zeman:2004:RC,Zeman:MSMSE:2007}. The CLT
approaches are, on the other hand, easy to implement and provide a
reasonable approximation of the in-plane elastic moduli. However,
since this class of models approximates the composite as a coupling of
serial and parallel laminates stacked to resemble the actual geometry,
it becomes inadequate when predicting the out-of-plane
response. Therefore, a procedure offering a reasonable compromise
between the accuracy of FEM-based modeling and simplicity of
traditional CLT methods is still on demand.

In the last decade, effective media theories, widely used in continuum
micromechanics~\cite{Bohm:2005:SIBA}, have been recognized as an
attractive alternative to CLT-based methods. Such an approach was
pioneered by~\citeasnoun{Gommers:1998:MTMATC}
and~\citeasnoun{Huysmans:1998:PIA}, who modeled knitted composites as
an assembly of spherical fibers in an isotropic matrix and used the
Mori-Tanaka~(\MT) method~\cite{Mori:1973:MTM} to evaluate the overall
response. Further advancements in the field include the Transformation
Field Analysis approach due to~\citeasnoun{Bahei-El-Din:2004:MMD} and
the work of ~\citeasnoun{Barbero:2005:MFRC} in the framework of the
theory of periodic eigenstrains. All these studies report good
correspondence with experimental data with an error comparable to
experimental scatter. Moreover, with regard to imperfect \delete{\CC}
textiles, the Mori-Tanaka method appears particularly useful as it
allows, through the application of orientation averaging techniques,
see e.g ~\cite[and references
  therein]{Yushanov:1998:STCM,Gommers:1998:MTMATC,Schjodt-Thomsen:2001:MT,Duschlbauer:2003:MTB,Jing:2003:MESF,Doghri:2006:MIE},
for a direct introduction of imperfections in the fiber-tow path
represented here by histograms of distribution of the fiber-tow
orientation angles. It is worth noting that such histograms, when
constructed for all plies in the laminate, also reflect, at least to
some extent, \replace{their mutual}{tow path imperfections due to
  inter-layer} shift typical for real material systems displayed in
\Fref{F:EPUC}(a),(b).

A successful application of the \MT~method for the prediction of the
effective material parameters of textile \delete{\CC} composites
including the above knowledge of the actual microstructure requires,
however, completion of the following tasks:

\begin{itemize}

\item Quantification of the real micro (meso) structure through a
  detailed evaluation of images of real material samples. This part of
  the analysis is briefly addressed in \Sref{sec:micro} \new{for a
    \CC~composite specimen}.

\item The basic geometrical information are then used to construct an
  idealized three-dimensional periodic unit cell exploiting the
  geometrical model proposed by~\citeasnoun{Kuhn:1999:MPWCG}. Such a
  unit cell serves as a point of departure for a subsequent
  application of the \MT~method. Review of the model together with
  essential steps of the FEM based simulations using the first-order
  homogenization technique is provided in \Sref{sec:puc}.

\item Formulation of the Mori-Tanaka method is then presented in
  \Sref{sec:mt} with emphasis given to the symmetry of the overall
  material stiffness matrix and capturing interaction between
  individual tows. It is shown that special care is required when
  replacing the actual fiber tow by an equivalent ellipsoidal
  inclusion. In the present formulation, the shape of the equivalent
  ellipsoid is thus treated as an internal parameter of the method
  determined by matching the \MT~estimates with the results derived in
  \Sref{sec:puc}.

\item Two possible approaches to the calibration of the internal
  parameters of the method are considered. In the first variant, the
  optimal ellipsoidal shape is found by matching directly the results
  of the finite element simulations, executed on a ``training'' set
  reflecting the in-situ determined scatter of geometrical
  parameters. It is worth noting that such procedure allows us to
  introduce possible deviations from the ``ideal'' (average) geometry
  of the textile structure presented in \Sref{sec:micro}. The second
  approach employs the finite element data at hand to propose a simple
  relation between basic parameters of the textile composite and
  parameters of the optimal ellipsoid. \Sref{sec:mt-optimal} is
  concluded by verification and validation of the developed
  heuristics.

\item Once the shape of the ellipsoid is calibrated, the orientation
  averaging can be used in conjunction with the histograms of the
  fiber-tow orientation angle. This final step leading to estimates of
  the overall elastic response of imperfect \CC~textile composites is
  examined in \Sref{sec:mt-real}. \Sref{sec:con} then summarizes the
  final results and offers possible extensions particularly with
  account to intrinsic porosity of these material systems.

\end{itemize}

In the following text, the Voigt representation of symmetric tensorial
quantities is systematically employed,
e.g.~\cite{Bittnar:1996:NMM}. In particular, $a$, $\mtrx{a}$ and
$\mtrx{A}$ denote a scalar value, a vector or a matrix representation
of a second-order tensor and a matrix representation of a fourth-order
tensor, respectively. Other symbols and abbreviations are introduced
in the text as needed.

\section{Microstructure evaluation}\label{sec:micro}
%%%%%%%%%%%%%%%%%%%%%%%%%%%%%%%%%%%

\begin{figure}[ht]
\centerline{\figName{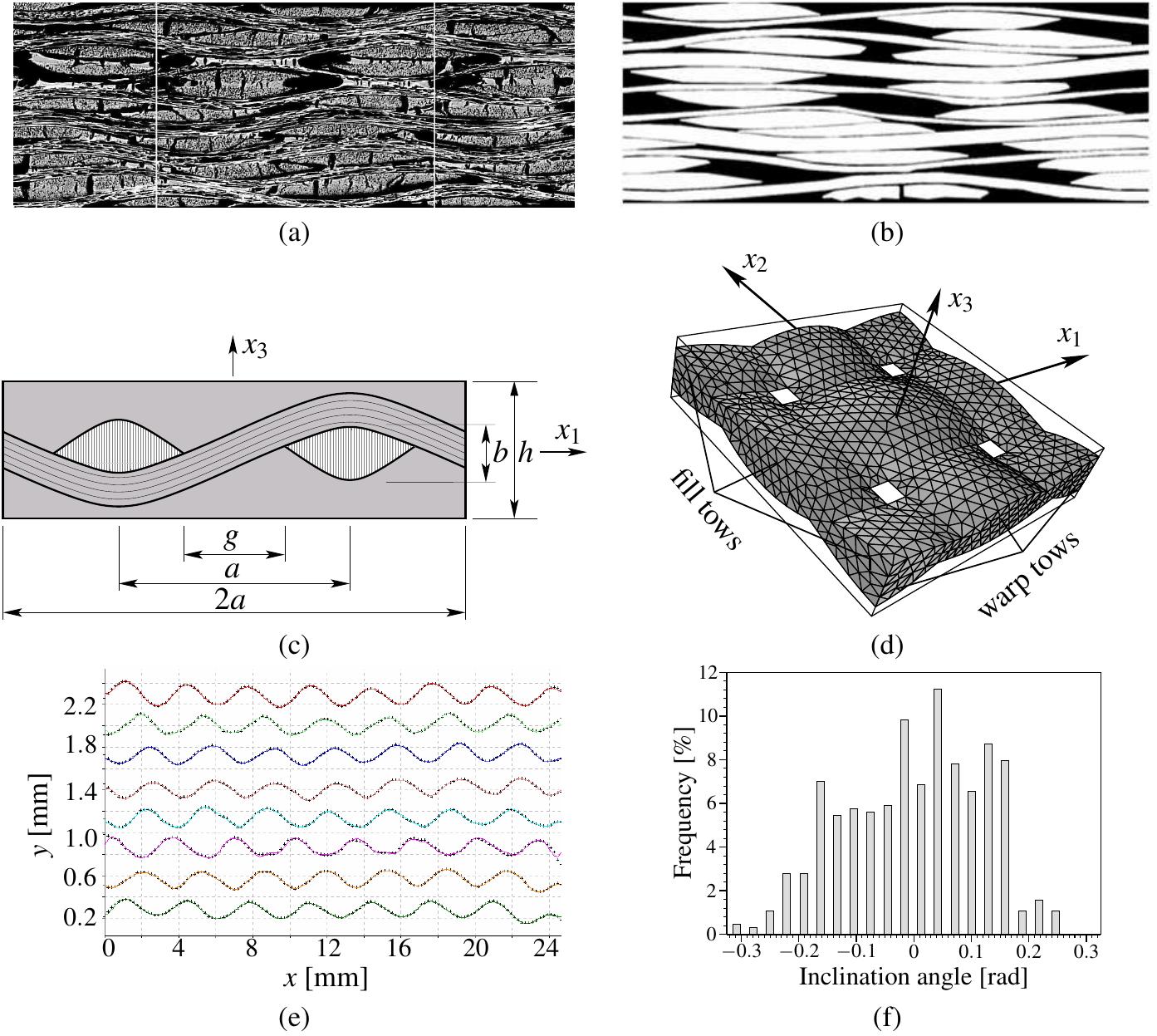}}
\caption{Equivalent periodic unit cells; (a) color image of real
  composite sample, (b) binary image, (c) cross-section of an equivalent
  periodic unit cell, (d) three-dimensional view, (e) approximation of
  centerlines~\cite{Vopicka:2004:PGV}, (f) distribution of inclination angles.}
\label{F:EPUC}
\end{figure}

As already mentioned in the introductory part, obtaining reliable
predictions of the effective mechanical properties of \delete{\CC}
textile composites requires a thorough analysis of their actual
microstructure. \Fref{F:EPUC}(a) shows a particular
\new{\CC}~composite laminate consisting of eight layers of carbon
fabric Hexcel G~1169 bonded to a carbon matrix. A total of twenty such
specimens having dimensions of $25 \times 2.5 \times 2.5$~mm were
fixed into the epoxy resin and after curing subjected to final surface
grounding and polishing using standard metallographic techniques to
produce specimens suitable for the subsequent image analysis.

While image analysis software \code{LUCIA G}$^{\mbox{\textregistered}}$ might
be used directly to process the actual image of the specimen in
\Fref{F:EPUC}(a), it proves more advantageous, owing to a low color
contrast of the carbon reinforcement and carbon matrix, to collect the
necessary geometrical information from its binary counterpart plotted
in \Fref{F:EPUC}(b). Several such sections taken from various
locations of the laminated plates were examined to obtain basic
statistics of various parameters including segment dimensions, fiber
tow thickness, shape of the fiber tow cross-section, etc. The
resulting values are stored in the second column of \Tref{T:geom}. The
averages of basic geometrical data were finally used to construct an
equivalent or rather an ideal periodic unit cell (EPUC) appearing in
\Fref{F:EPUC}(c,d) employing the description due
to~\citeasnoun{Kuhn:1999:MPWCG}. The three-dimensional geometric model
is defined by four parameters: the tow wavelength $2a$, the tow height
$b$, tow spacing $g$ and the layer thickness $h$,
cf. \Fref{F:EPUC}(c). 

\begin{table}[hb]
\caption{Quantification of microstructural parameters}
\label{T:geom}
\centering
\begin{tabular}{lrrr}
\br
          & Carbon/Carbon              & E-glass/Vinylester 
& E-glass/Epoxy \\
Parameter & \citeasnoun{Tomkova:2004a} & \citeasnoun{Scida:1999:MM3D}
& \citeasnoun{Kollegal:2000:SMP} \\
\mr
$a$~[$\mu$m]       & 2,250 $\pm$ 155 & 1,200 & 620 \\
$h$~[$\mu$m]       & 300  $\pm$ 50   & $\times$ & $\times$ \\
$b$~[$\mu$m]       & 150  $\pm$ 20   & 50 & 100 \\
$g$~[$\mu$m]       & 400  $\pm$ 105  & 20 & 20 \\
$c_{\rm tow}$ [$\%$] & 53.2 $\pm$ 1.8  & 79.8  & 69.7 \\
\br
\end{tabular}
\end{table}

Note, however, that the real composite shows a number of imperfections
which certainly should not be completely disregarded. It will be seen
later in \Sref{sec:mt} that the nonuniform waviness and to some extent
\replace{the}{also the fiber inclinations due to production-related}
mutual shift of individual layers clearly visible in \Fref{F:EPUC}(b)
can be accounted for by utilizing histograms of inclination angles
derived from centerlines of individual fiber tows, see
\Fref{F:EPUC}(e,f) and~\cite{Vopicka:2004:PGV} for more details. The
idealized geometry in \Fref{F:EPUC}(c) assumes, nevertheless, the
centerlines of the warp and fill systems of tows in a simple
trigonometric form~\cite{Kuhn:1999:MPWCG}
\begin{equation}\label{eq:centerline}
c( x ) = \frac{b}{2} \sin \left( \frac{\pi x}{a} \right).
\end{equation}

\section{Periodic unit cell analysis}\label{sec:puc}
%%%%%%%%%%%%%%%%%%%%%%%%%%%%%%%%%%%%%%%%%%%%%%%%%%%%
Having quantified the real microstructure, the resulting EPUC can be
readily employed to provide FEM estimates of the required effective
moduli. This particular step of the proposed analysis scheme will now
be briefly reviewed.

\begin{figure}[b]
\centerline{\figName{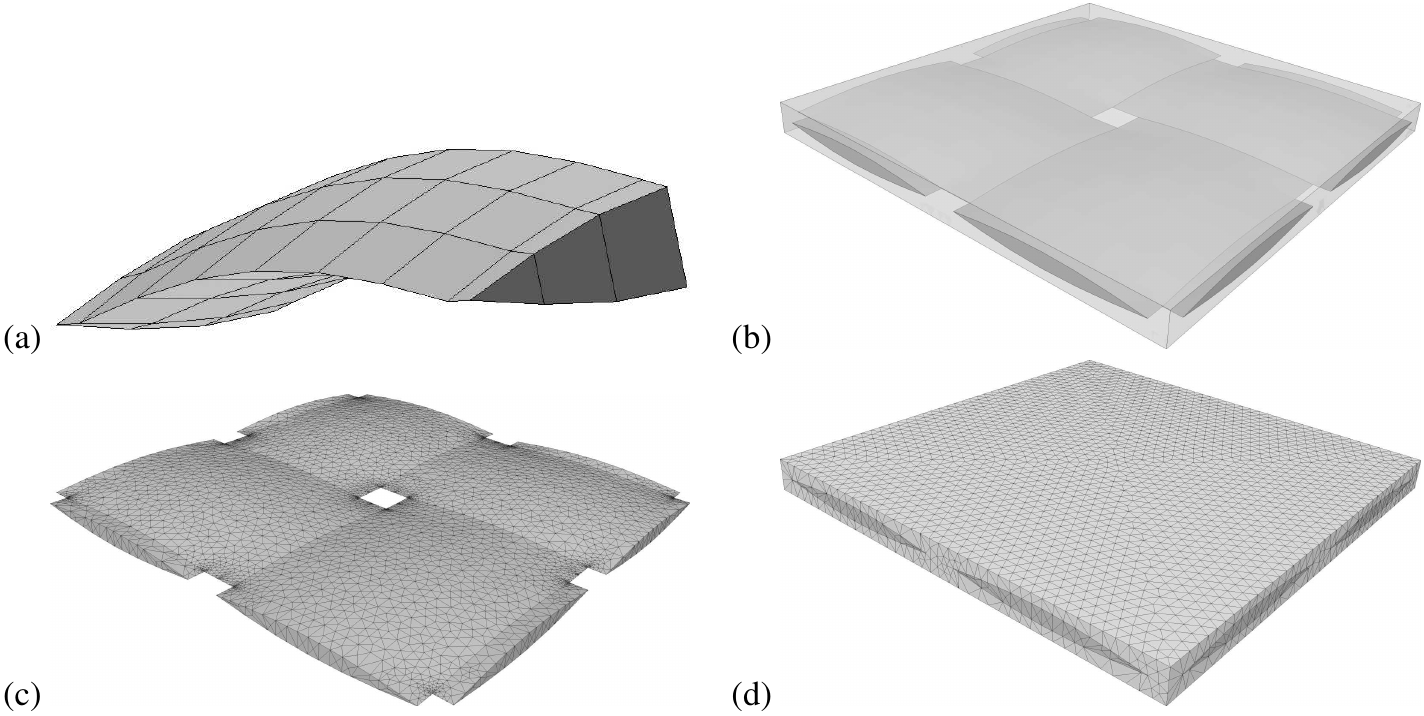}}
\caption{Finite element mesh generation;
	(a)~CAD\delete{E} model of primitive volume, 
  	(b)~CAD model of PUC, 
	(c)~FEM mesh of fiber tows, 
	(d)~FEM mesh of PUC.}
\label{fig:mesh_generation}
\end{figure}

To that end, consider an EPUC in \Fref{F:EPUC}(d) with the local
coordinate system defined such that the local $x_1^\ell$ axis is
aligned with the fiber tow direction. Definitely the most tedious step
in the entire analysis is preparation of a three-dimensional finite
element mesh complying with the periodic boundary conditions (the same
positions of the element nodes on the opposite faces of the
cell). Here, the elements of CAD operations combined with volumetric
modeling capabilities of \code{ANSYS}$^{\mbox{\textregistered}}$ package
are used to generate the finite element mesh employing the mapped
meshing technique discussed by~\citeasnoun{Wentorf:1999:AMCW}
and~\citeasnoun{Matous:2007:MMSP}.

In order to ensure symmetry of the resulting FEM mesh, a primitive
block of the tow shown in~\Fref{fig:mesh_generation}(a) is modeled
first. Next, using mirroring, copying and merging operations, the
whole volume of one reinforcement layer is generated. Finally, the
volume corresponding to the matrix phase is generated by subtracting
the body of reinforcements from the matrix as depicted
in~\Fref{fig:mesh_generation}(b). To reflect the required periodicity
only one of the two opposite faces is meshed using the advancing front
technique and then copied to the associated one. At last, the
tetrahedral elements corresponding to tows and matrix are generated
based on the data created in the previous steps, leading to finite
element meshes shown in \Fref{fig:mesh_generation}(c,d).

The further numerical treatment now proceeds as follows,
cf.~\cite{Michel:1999:EPC}. Suppose that the periodic unit cell in
\Fref{fig:mesh_generation}(d) is loaded by a macroscopic strain vector
$\mtrx{E}$. In view of the assumed microstructure periodicity, the
local displacement field $\mtrx{u}$ then admits the following
decomposition
\begin{equation}
\mtrx{u}(\mtrx{x}) = \mtrx{X}(\mtrx{x}) \mtrx{E} 
+
\mtrx{u}^{*}(\mtrx{x}),  
\end{equation}
where $\mtrx{u}^{*}$ represents a periodic fluctuation of $\mtrx{u}$
due to the presence of heterogeneities and matrix $\mtrx{X}$ stores
the coordinates of $\mtrx{x}$. The local strain then assumes the form
\begin{equation}\label{eq:localeps} 
\mtrx{\strain}(\mtrx{x}) = \mtrx{E} +
\mtrx{\strain}^*(\mtrx{x}), 
\end{equation} 
where the fluctuating part $\mtrx{\strain}^*$ vanishes upon the volume
averaging. Next, introducing \Eref{eq:localeps} into the principle of
virtual work (the Hill-Mandel lemma) yields
\begin{equation}\label{eq5:hlemma}
\avgs{\delta\mtrx{\strain}\trn(\mtrx{x}) \mtrx{\stress}( \mtrx{x} )} 
= \avgs{ \delta{\mtrx{\strain}^\ell}\trn(\mtrx{x}) \mtrx{\stress}^\ell( \mtrx{x} ) }
= \avgs{\delta{\mtrx{\strain}^{*\ell}}\trn(\mtrx{x}) \mtrx{\stress}^\ell( \mtrx{x} ) }
= 0,
\end{equation}
where $\avgs{~}$ stands for the volumetric averaging with respect
to the PUC and $\cdot^\ell$ is used to denote a quantity in the local
coordinate system. The local stress field then reads
\begin{equation}\label{eq5:localstress}
\mtrx{\stress}^\ell(\mtrx{x}) = \mtrx{L}^\ell(\mtrx{x})
\left(  \mtrx{E}^\ell + \mtrx{\strain}^{*\ell} (\mtrx{x})
\right),
\end{equation}
where $\mtrx{L}^\ell$ is the material stiffness matrix. Relating the
strains in the local and global coordinate systems by the well-known
relations $\mtrx{E}^\ell = \mtrx{T}_\strain \mtrx{E}$,
$\mtrx{\strain}^\ell = \mtrx{T}_\strain \mtrx{\strain}$, see
e.g.~\cite{Bittnar:1996:NMM}, and inserting \Eref{eq5:localstress}
into \Eref{eq5:hlemma} yields the stationarity conditions in the form
\begin{equation}
\left\langle \delta {\mtrx{\strain}^*}\trn( \mtrx{x} ) 
\mtrx{T}_\strain\trn( \mtrx{x}) \left[ \mtrx{L}^\ell(\mtrx{x})
 \mtrx{T}_\strain( \mtrx{x} ) \left( \mtrx{E} + \mtrx{\strain}^*(
  \mtrx{x}) \right) \right] \right\rangle = 0,
\end{equation} 
to be satisfied for all kinematically admissible variations $\delta
\mtrx{\strain}^*$.

The homogenized stiffness matrix $\mtrx{L}^{\rm FEM}$ follows from
post-processing of the solution of six independent elasticity
problems, discretized using conforming FEM procedure,
see~\cite{Zeman:2003:ACM,Zeman:2004:RC} for further details. In
particular, each column of $\mtrx{L}^{\rm FEM}$ coincides with the
volume averages of local stress $\mtrx{\stress}$ resulting from a
macroscopic strain with one component set to one and with the
remaining entries equal to zero.

\section{Application of the Mori-Tanaka to woven composites}\label{sec:mt}
%%%%%%%%%%%%%%%%%%%%%%%%%%%%%%%%%%%%%%%%%%%%%%%%%%%%%%%%%%%

\subsection{Overall stiffness of composite with non-aligned inclusions}\label{sec:mt-theory}
%%%%%%%%%%%%%%%%%%%%%%%%%%%%%%%%%%%%%%%%%%%%%%%%%%%%%%%%%%%%%%%%%%%%%%%
Consider an $N$-phase composite with an isotropic matrix phase having
the stiffness matrix $\mtrx{L}_0$ and being reinforced with $(N-1)$
families of ellipsoidal heterogeneities. Each heterogeneity is
characterized by the stiffness matrix $\mtrx{L}_r$ and occupies a
volume $\Omega_r$. With reference to~\cite{Benveniste:1991:ODES}, the
Mori-Tanaka estimate of the overall stiffness matrix $\mtrx{L}^{\rm \MT}$ then
reads
\begin{equation}\label{eq:mt_orig}
\mtrx{L}^{\rm \MT} = 
\mtrx{L}_0 
+ 
\left(
 \sum_{r=1}^{N-1} c_r \left( \mtrx{L}_r - \mtrx{L}_0 \right)\mtrx{T}_r
\right)
\left(
c_0 \mtrx{I} 
+ 
\sum_{r=1}^{N-1} c_r\mtrx{T}_r
\right)^{-1},
\end{equation}
where $c_r$ denotes the volume fraction of the $r$-th phase. The
corresponding partial \replace{stress}{strain} concentration factor $\mtrx{T}_r$ has
the form
\begin{equation}\label{eq:Tr}
\mtrx{T}_r = \left( \mtrx{I} + \mtrx{P}_r \left( \mtrx{L}_r -
  \mtrx{L}_0 \right) \right)^{-1},
\end{equation}
where the $\mtrx{P}_r$ matrix is provided by
\begin{equation}
\mtrx{P}_r = \int_{\Omega_r} \mtrx{\Gamma}_0( \mtrx{x} - \mtrx{x}')
  \de \mtrx{x}'.
\end{equation}
Function $\mtrx{\Gamma}_0$ is related to Green's function of an
infinite medium with stiffness matrix $\mtrx{L}_0$~(see,
e.g.~\cite[Section 3.1]{Castenada:1995:ESD} for more details). It
follows from the celebrated work of~\citeasnoun{Eshelby:1957:SFI} that for
ellipsoidal inclusions, $\mtrx{P}_r$ is constant and can be evaluated
as
\begin{equation}\label{eq:Pr}
\mtrx{P}_r = \mtrx{S}_r \mtrx{L}_0^{-1},
\end{equation}
where $\mtrx{S}_r$ is the Eshelby matrix. When the matrix phase is
isotropic, explicit expressions for $\mtrx{S}_r$ can be found in,
e.g.~\cite{Eshelby:1957:SFI,Mura:1982:MDS}.

While the \MT~model has proved itself to be accurate for composites
reinforced either by randomly oriented or aligned inclusions with an
identical shape, in general case it may lead to a non-symmetric
stiffness matrix $\mtrx{L}^{\rm \MT}$, see
e.g.~\cite{Benveniste:1991:ODES,Ferrari:1991:AHL,Castenada:1995:ESD}
for the in-depth discussion. In this work, a simple re-formulation
proposed by~\citeasnoun{Schjodt-Thomsen:2001:MT} is employed to
preserve the overall symmetry of the stiffness matrix using the
orientation averaging. 

\begin{figure}[ht]
\centerline{\figName{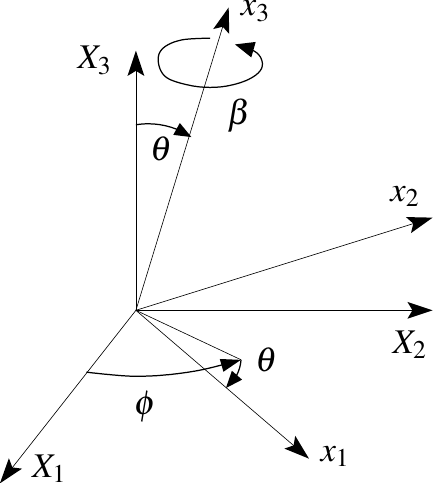}}
\caption{Definition of the Euler angles.}
\label{fig:euler}
\end{figure}

To this end, we \replace{limit our attention to}{approximate the
  material system under investigation as} a two-phase
\replace{material system}{composite} ($N=2$) consisting of an
isotropic matrix \replace{reinforced by identical
  heterogeneities with different orientations}{($r=0$) and with index $r=1$
  collectively denoting the reinforcing tow phase, composed of
  heterogeneities of identical shape but different
  orientations.\footnote{%
Therefore, the employed geometrical data reduce to the volume fraction
of one of the phases and appropriate quantification of orientation
distribution.}} Suppose for a moment \replace{that all}{aligned}
heterogeneities \delete{posses the same orientation}. Then, the
overall stiffness matrix is symmetric~\cite{Benveniste:1991:ODES} and
can be decomposed to
\begin{equation}
\mtrx{L}^{\rm \MT} = \mtrx{L}_0 + c_1
\left(\left( \mtrx{L}_1 - \mtrx{L}_0 \right)\mtrx{T}_1\right)
\left(( 1 - c_1 ) \mtrx{I} + c_1\mtrx{T}_1\right)^{-1} 
=  \mtrx{L}^{(0)} + c_1 \mtrx{L}^{(1)}.
\end{equation}
Notice that due to assumed isotropy of the matrix phase, the matrix
$\mtrx{L}^{(0)}$ is independent of the reference coordinate system
while $\mtrx{L}^{(1)}$ stores the orientation-dependent part.
Following~\cite{Schjodt-Thomsen:2001:MT}, the estimate of the overall
stiffness \new{of a composite reinforced with non-aligned
  heterogeneities} is provided by
\begin{equation}\label{Schjodt}
\mtrx{L}^{\rm \MT} \approx \mtrx{L}^{(0)} 
+
c_1 \oavg{\mtrx{L}^{(1)}}, 
\end{equation}
where the double brackets $\oavg{~}$ denote averaging over all
possible orientations. In particular, when the orientation of each
heterogeneity is parametrized in terms of the Euler angles $\phi,
\theta$ and $\beta$, see \Fref{fig:euler},\footnote{Note that
  so-called "$x_2$ convention" is used; i.e. a conversion into a new
  coordinates system follows three consecutive steps. First, the
  rotation of angle $\phi$ around the original $X_3$ axis is
  done. Then, the rotation of angle $\theta$ around the new $x_2$ axis
  is followed by the rotation of angle $\beta$ around the new $x_3$
  axis to finish the conversion.}
the orientation-dependent part can be expressed as
\begin{equation}
\mtrx{L}^{(1)}( \theta, \phi, \beta ) =
\mtrx{T}_\strain\trn( \theta, \phi, \beta ) \mtrx{L}^{(1)}( 0, 0, 0 )
\mtrx{T}_\strain( \theta, \phi, \beta ),
\end{equation}
where the explicit expression of the transformation matrix can be
found in e.g.~\cite{Schjodt-Thomsen:2001:MT} and~\cite[Appendix
  A]{Zeman:2003:ACM}.

The orientation average then follows from
\begin{equation}\label{orient_average}
\oavg{\mtrx{L}^{(1)}} = \int_0^{2\pi} \int_0^{2\pi} \int_{0}^{\pi} 
\, \mtrx{L}^{(1)}( \theta, \phi, \beta ) g( \theta, \phi, \beta ) \de \theta \de
\phi \de \beta,
\end{equation}
with $g( \theta, \phi, \beta )$ denoting the joint probability density
describing the distribution of individual angles.

\subsection{Application to plain weave composites with ideal geometry}\label{sec:mt-ideal}
%%%%%%%%%%%%%%%%%%%%%%%%%%%%%%%%%%%%%%%%%%%%%%%%%%%%%%%%%%%%%%%%%%%%%%

To examine the theoretical formulation presented in the previous
Section, consider again an ideal plain weave textile composite already
studied in~\Sref{sec:puc}. In this particular case, the joint
probability density function $g(\theta,\phi,\beta)$ results from the
harmonic shape of the centerline, recall
\Eref{eq:centerline}. Applying the change of variable
formula~\cite[Section 33.9]{Rektorys:1994:SOM}, we obtain after some
algebra the expression of the probability density in the form
$$
g( \theta, \phi, \beta ) = \left\{
\begin{array}{cl}
\displaystyle
\frac{2a}{\pi}\frac{1+\tan^2(\theta)}{\sqrt{b^2\pi^2-4a^2\tan^2(\theta)}}
& \mbox{if } \phi = 0, \beta = 0 \mbox{ and }
-\alpha \leq \theta \leq \alpha, \\
0 & \mbox{otherwise,}
\end{array}
\right.
$$
where
$$
\alpha = \arctan\left( \frac{b\pi}{2a} \right).
$$
\Eref{orient_average} then becomes
\begin{equation}\label{eq:warp}
\oavg{\mtrx{L}^{(1,{\rm warp})}} 
= 
\frac{2a}{\pi}
\int_{-\alpha}^{\alpha} 
\frac{1+\tan^2(\theta)}{\sqrt{b^2\pi^2-4a^2\tan^2(\theta)}}
\mtrx{L}^{(1)}( 0, \theta, 0 ) 
\de \theta, 
\end{equation}
and similarly for the fill system we get
\begin{equation}
\oavg{\mtrx{L}^{(1,{\rm fill})}} 
= 
\frac{2a}{\pi}
\int_{\pi/2-\alpha}^{\pi/2+\alpha}  
\frac{1+\tan^2(\theta)}{\sqrt{b^2\pi^2-4a^2\tan^2(\theta)}}
\mtrx{L}^{(1)}( \frac{\pi}{2}, \theta, 0 )
\de \theta.
\end{equation}
Following \Eref{Schjodt}, the resulting homogenized stiffness matrix
of a plain weave composite then reads
\begin{equation}\label{eq:woven_mt}
\mtrx{L}^{\rm \MT}(c_1, a, b, \mtrx{L}_0, \mtrx{L}_1,\mtrx{S}_1) 
= 
\mtrx{L}^{(0)} 
+ 
\frac{c_1}{2} \left( 
\oavg{\mtrx{L}^{(1,{\rm fill})}} 
+ 
\oavg{\mtrx{L}^{(1,{\rm warp})}}
\right)
,
\end{equation}
where the averages of the basic geometrical parameters $a, b$ and the
material parameters of the two-phase composite including the volume
fraction of individual phases are assumed to be known quantities. The
matrices $\oavg{\mtrx{L}^{(1,{\rm warp})}}$ and
$\oavg{\mtrx{L}^{(1,{\rm fill})}}$ implicitly depend, however, on the
Eshelby matrix $\mtrx{S}_1$, which is yet to be determined.

\begin{table}[b]
\caption{Material parameters of individual phases in local coordinate
  system}
\label{T:phases}
\centering
\begin{tabular}{lrrr}
\br
& Carbon/Carbon & E-glass/\replace{vinyelster}{Vinylester} & E-glass/\replace{e}{Epoxy} \\
&               & \citeasnoun{Barbero:2005:MFRC} & \citeasnoun{Barbero:2005:MFRC} \\
\mr
\emph{Matrix} \\
$E$~[GPa] & 30   & 3.4  & 3.12 \\
$\nu$     & 0.19 & 0.35 & 0.38 \\
\emph{Fiber tow} \\
$E_A$~[GPa] & 210  & 58.397 & 51.352 \\
$G_A$~[GPa] &  86  &  8.465 &  5.342 \\
$E_T$~[GPa] &  72  & 20.865 & 15.040 \\
$G_T$~[GPa] & 27.7 &  7.527 &  5.342 \\
$\nu_A$     & 0.27 &  0.241 &  0.262 \\
\br
\end{tabular}
\end{table}

To take advantage of the \new{closed-form} Eshelby
solution~\cite{Eshelby:1957:SFI}, it will be assumed that the actual
shape of the fiber tow can be well represented by an equivalent
ellipsoid with semi axes $\sa_1 \geq \sa_2 \geq \sa_3 > 0$. Then, the
accuracy of the \MT~method is governed by a proper choice of the
semi-axes as exemplified by the following case study. In particular,
three representations \replace{are}{is} considered: (i)~a spherical
shape $(\sa_1=\sa_2=\sa_3)$, (ii)~a cylindrical shape
$(\sa_1\rightarrow \infty,\sa_2=\sa_3)$ and (iii)~an
ellipsoid~$(\sa_1=1,\sa_2=0.5,\sa_3=0.1)$. The unit cell with average
geometrical parameters appearing in the second column of \Tref{T:geom}
and the constituent properties stored in the second column of
\Tref{T:phases}\new{, i.e. \CC~composite system,} are considered. Note
that in order to achieve the maximum phase stiffness contrast, the tow
parameters shown in~\Tref{T:phases} correspond to the pure carbon
fibers. The corresponding homogenized stiffness matrix entries are
stored in \Tref{T:hmg_stiff} together with the \FEM~data.
\begin{table}[ht]
\centering
\caption{Homogenized stiffness matrix entries (\CC~system)}
\label{T:hmg_stiff}
\begin{tabular}{lrrrrrr}
\br
Method & 
$L_{11}$ & $L_{12}$ & $L_{13}$ &
$L_{33}$ & $L_{44}$ & $L_{66}$ \\ 
& [GPa] & [GPa] & [GPa] & [GPa] & [GPa] & [GPa] \\
\mr
FEM~(Ideal geometry) & 
89.94 & 16.55 & 14.06 & 48.67 & 20.26 & 41.53 \\
\MT~(Spherical inclusion) & 
64.73 & 14.57 & 14.85 & 54.56 & 24.17 & 28.71 \\
\MT~(Cylindrical inclusion) & 
103.6 & 16.15 & 15.55 & 52.74 & 23.99 & 28.75 \\
\MT~(Ellipsoidal inclusion) & 
88.35 & 16.85 & 14.94 & 50.24 & 21.54 & 41.19 \\
\br
\end{tabular}
\end{table}

Clearly, comparing the \MT~estimates with FEM based results allows us
to draw the following two conclusions: (i)~the Mori-Tanaka method
appears as a reliable alternative to the first-order periodic
homogenization based on the finite element method, (ii)~the choice of
the Eshelby matrix can hardly be made arbitrarily.  \new{It should be
  noted that the illustrative results correspond to volume fraction
  $c_1 \approx 50\%$, for which the accuracy of the Mori-Tanaka method
  typically deteriorates. Therefore, the optimal ellipsoid not only
  accounts for the tow geometry, but also for approximately captures
  tow interactions due to non-dilute volume fractions of the
  reinforcing phase.}

\section{Optimal shape of equivalent ellipsoid}\label{sec:mt-optimal}
%%%%%%%%%%%%%%%%%%%%%%%%%%%%%%%%%%%%%%%%%%%%%%%%%%%%

\subsection{FEM-based calibration}
%%%%%%%%%%%%%%%%%%%%%%%%%%%%%%%%%%%
The essential goal now becomes to find the optimal shape of the
ellipsoid by matching the FEM results with the \MT~predictions
\new{for \CC~material system}. To take into account the observed
uncertainties in the textile geometry, a collection of PUCs, rather
than a single one, is used for the calibration. To generate such a set
we exploit the scale-invariance of the first order homogenization and
set $a=1$. The remaining parameters were generated using the Latin
Hypercube Sampling method~\cite{Iman:1980:SSS}, assuming uniformly
distributed random variables with the statistics stored in the second
column of~\Tref{T:geom}. Twenty such unit cells were generated and
subject to the FEM-based homogenization procedure, yielding the
homogenized stiffnesses listed in \Tref{T:Lset}.

\begin{table}[ht]
\caption{Summary of homogenized effective properties of training set~\new{(\CC~system)}}
\label{T:Lset}
\centering
\begin{tabular}{lrrrrrr}
\br
Statistics &
$L^{\rm FEM}_{11}$ & $L^{\rm FEM}_{12}$ & $L^{\rm FEM}_{13}$ & 
$L^{\rm FEM}_{33}$ & $L^{\rm FEM}_{44}$ & $L^{\rm FEM}_{66}$ \\ 
& [GPa] & [GPa] & [GPa] & [GPa] & [GPa] & [GPa] \\
\mr
Average            & 86.96 & 16.00 & 13.65 & 47.73 & 19.78 & 39.13 \\
Standard deviation &  2.29 &  0.40 &  0.30 &  0.68 &  0.34 & 1.87 \\
\new{Optimized \MT~(ideal geometry)} & 
88.81 & 16.13 & 13.89 & 47.35 & 20.17 & 40.40 \\
\new{Optimized \MT~(training set)} &
88.23 & 16.78 & 14.94 & 49.09 & 20.09 & 40.26 \\
\br
\end{tabular}
\end{table}

\begin{figure}
\centerline{\figName{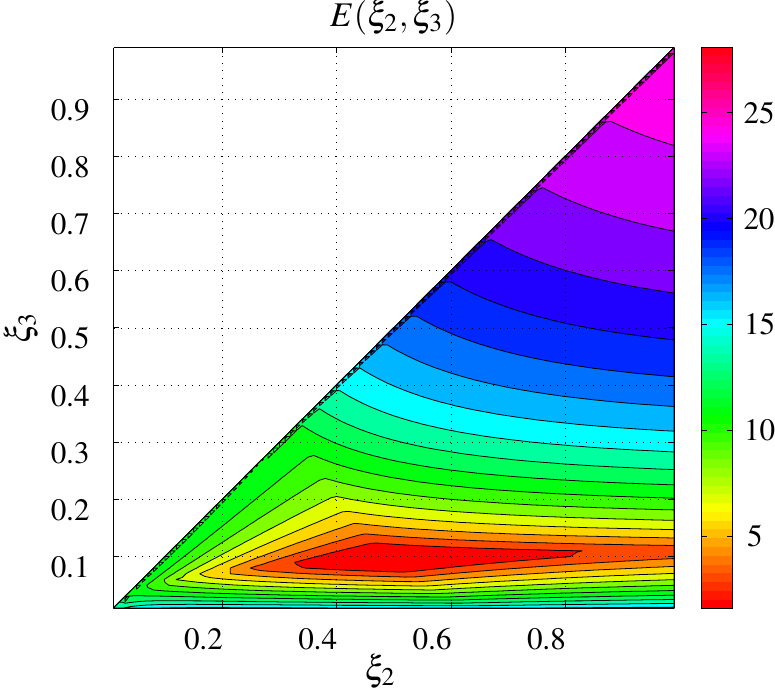}}
\caption{Objective function for a single PUC.}
\label{F:objective}
\end{figure}

Since the Eshelby matrix depends on the mutual ratio of the ellipsoid
semi-axes only, it is possible to set $\sa_1 = 1$, which leaves us
with only two parameters undetermined. To characterize the discrepancy
between the FEM and \MT~solution, the following error measure is
introduced
\begin{equation}\label{eq:objfunc_single}
E( \mtrx{L}^{\rm FEM}, \mtrx{L}^{\rm \MT} ) 
= 
\max_{i,j=1,\ldots,6}
\left|
L^{\rm FEM}_{ij} - L^{\rm \MT}_{ij}
\right|.
\end{equation}
When $n$ PUCs are considered, the objective function assumes the form:
\begin{equation}\label{eq:obj_func}
F( \sa_2, \sa_3 ) = \sqrt{\sum_{i=1}^{n} E^2\left( \mtrx{L}_{(i)}^{\rm FEM},
\mtrx{L}_{(i)}^{\rm \MT}( \sa_2, \sa_3 )\right)},
\end{equation}
where the superscript $i$ represents the $i$-th member in the training
set. The optimal shape characterized by $\sa_2^*$ and $\sa_3^*$ can be
then found from the minimization procedure
\begin{equation}\label{eq4:obj_func}
( \sa^*_2, \sa^*_3 ) 
\in 
\arg \min_{0 \leq \sa_2 \leq 1, \sa_2 \leq \sa_3 \leq 1} 
F( \sa_2, \sa_3 ).
\end{equation}
A graphical representation of the objective function assuming a single
(average) periodic unit cell is plotted in \Fref{F:objective} for the
sake of illustration.

Note that the explicit expression of the Eshelby matrix is not
available in this particular case, which essentially precludes the use
of classical gradient-based optimization algorithms. The stochastic
optimization methods, on the other hand, appear to be a more
appropriate choice. The particular algorithm, based on the surrogate
function model combined with evolutionary algorithm adopted in the
present study is briefly described in \ref{sec:app}.

The solution of the optimization problem then yields the optimal
values of semi-axes $\sa_2^* \doteq 0.486$ and $\sa_3^* \doteq 0.092$
with the optimal value $F^* \doteq 1.6$. It is worth noting that the
optimum compares rather well with the case when the minimization is
performed with respect to the ideal unit cell only, for which $E^*
\doteq 1.3$\new{, see also \Tref{T:Lset} for a comparison in terms of
  stiffness matrix entries}. This confirms the predictive capabilities
of the \MT~approach, at least in the range of addressed geometrical
variations.

\subsection{Heuristic calibration}
%%%%%%%%%%%%%%%%%%%%%%%%%%%%%%%%%%%

Although the advocated \MT~approach seems to offer an efficient way to
the prediction of the homogenized properties, especially when handling
composites with random tow imperfections, cf.~\Sref{sec:mt-real}, it
still requires the reference \FEM~simulations to tune the Eshelby
matrix. Therefore, it appears advantageous to establish a heuristic
link between the \EPUC~parameters and optimal ellipsoidal shape. In
previous works~\cite{Gommers:1998:MTMATC,Huysmans:1998:PIA}, such a
relation was derived from the local centerline curvature and
calibrated using selected elastic constants. The current framework, on
the other hand, offers a possibility to systematically use information
contained in the previously generated set of \EPUC s.

In the first step of the analysis, the optimization procedure is
executed independently for each \EPUC\ (i.e. with objective function
\eref{eq:objfunc_single}), yielding a set of optimal parameters $\{
\sa^*_{2(i)}, \sa^*_{3(i)} \}, i = 1,2, \ldots, 21$. Subjecting the
results to correlation analysis, see
e.g.~\cite[Section~34.5]{Rektorys:1994:SOM}, reveals that the
$\sa^*_2$ parameter is strongly correlated with $g/a$ ratio~(with the
coefficient of correlation equal to $\approx -0.8$), while it is
almost independent of $b/a$ value. An analogous trend can be observed
between $\sa^*_3$ and $b/a$ parameter. Such results authorize us to
postulate a simple linear relation between the optimal ellipsoid shape
and \EPUC~parameters. The optimal fit, now determined using objective
function~\eref{eq4:obj_func}, finally leads to a semi-empirical formula
\begin{equation}\label{eq:Eshelby_heuristics}
\sa_2^* \approx 1 - \frac{3g}{a}, 
\sa_3^* \approx \frac{1}{10} - \frac{b}{3a}.
\end{equation}

No doubt, such heuristics still builds on a representative finite
\new{element simulations} and as such requires to be verified
and validated against independent data. In the current work, we
examine two plain weave composite systems thoroughly analyzed
in~\cite{Barbero:2005:MFRC,Barbero:2006:FEM}:
(i)~E-glass/\replace{vinyelster}{Vinylester}
composite~\cite{Scida:1999:MM3D}, (ii)~E-glass/\replace{e}{E}poxy
material system~\cite{Kollegal:2000:SMP}. The corresponding
geometrical data are stored in \Tref{T:geom}, while the material
\replace{data}{parameters} of individual constituents are available in
\Tref{T:phases}. It is worth noting that the considered material
systems offer a considerably different tow volume fractions and
elastic constants of individual phases than in the calibration
step. Moreover, to keep the validation objective, the comparison will
now be based on orthotropic engineering moduli~(see,
e.g.~\cite{Bittnar:1996:NMM}) rather than stiffness matrix entries.

\begin{table}[b]
\caption{Verification and validation of Mori-Tanaka against Periodic
  media method due to~\citeasnoun{Barbero:2005:MFRC}, Finite
  element simulation simulation from~\cite{Barbero:2006:FEM}
  and experimental data from~\cite{Scida:1999:MM3D,Kollegal:2000:SMP}.}
\label{T:validation}
\centering
\begin{tabular}{lrrrrrrrr}
\br
& \multicolumn{4}{c}{E-glass/\replace{v}{V}inylester} & \multicolumn{3}{c}{E-glass/\replace{e}{E}poxy} \\
& Experiment & \PMM & \FEM & \MT & Experiment & \PMM & \MT \\
\mr
$E_{11}=E_{22}$~[GPa] & 24.8 $\pm$ 1.1 & 25.1   & 24.5 & 25.8 &
19.29    & 18.9 & 19.2 \\
$E_{33}$~[GPa]       & 8.5  $\pm$ 2.6 & 10.5   & 10.3 & 12.4 &
$\times$ & 8.74 & 8.83 \\
$G_{23}=G_{13}$~[GPa] & 4.2 $\pm$ 0.7  & 2.91   & 3.16 & 4.08 &
$\times$ & 2.57 & 2.92 \\ 
$G_{12}$~[GPa]       & 6.5 $\pm$ 0.8  & 4.37   & 5.52 & 6.44 &
3.18     & 3.07 & 3.85 \\
$\nu_{23} = \nu_{13}$ & 0.28 $\pm$ 0.07 & 0.34  & 0.38 & 0.38 &
$\times$ & 0.44 & 0.46 \\
$\nu_{12}$           & 0.1 $\pm$ 0.01 &  0.12 & 0.13 & 0.14 &
0.2      & 0.13 & 0.13 \\
\br
\end{tabular}
\end{table}

For the E-glass/\replace{vinyelster}{Vinylester} composite, performance of the \MT~method is
compared with the Periodic Microstructure Model~(\PMM )~(an
alternative micromechanics-based method based on a detailed
geometrical model due to \citeasnoun{Barbero:2005:MFRC}), independent
finite element study in \code{ANSYS}$^{\mbox{\textregistered}}$ and
experimental data. Results of the comparison, reported in
\Tref{T:validation}, demonstrate a reasonable match between the
\MT~predictions and remaining values. Although the accuracy of the
Young moduli is somewhat inferior with respect to detailed models, the
shear behavior is predicted very well and also the values of the
Poisson ratios are consistent with the results of alternative
numerical approaches.

Similar conclusions can be made for the E-glass/\replace{e}{E}poxy textile system,
see \Tref{T:validation}, where even closer match between the detailed
numerical model can be observed. In overall, the presented data
provide an evidence that the heuristic
relation~\eref{eq:Eshelby_heuristics} leads to reasonably accurate
estimates of the homogenized elastic properties.

\section{Application to real geometry of \CC~composite system}\label{sec:mt-real}

\begin{figure}[ht]
\centerline{\figName{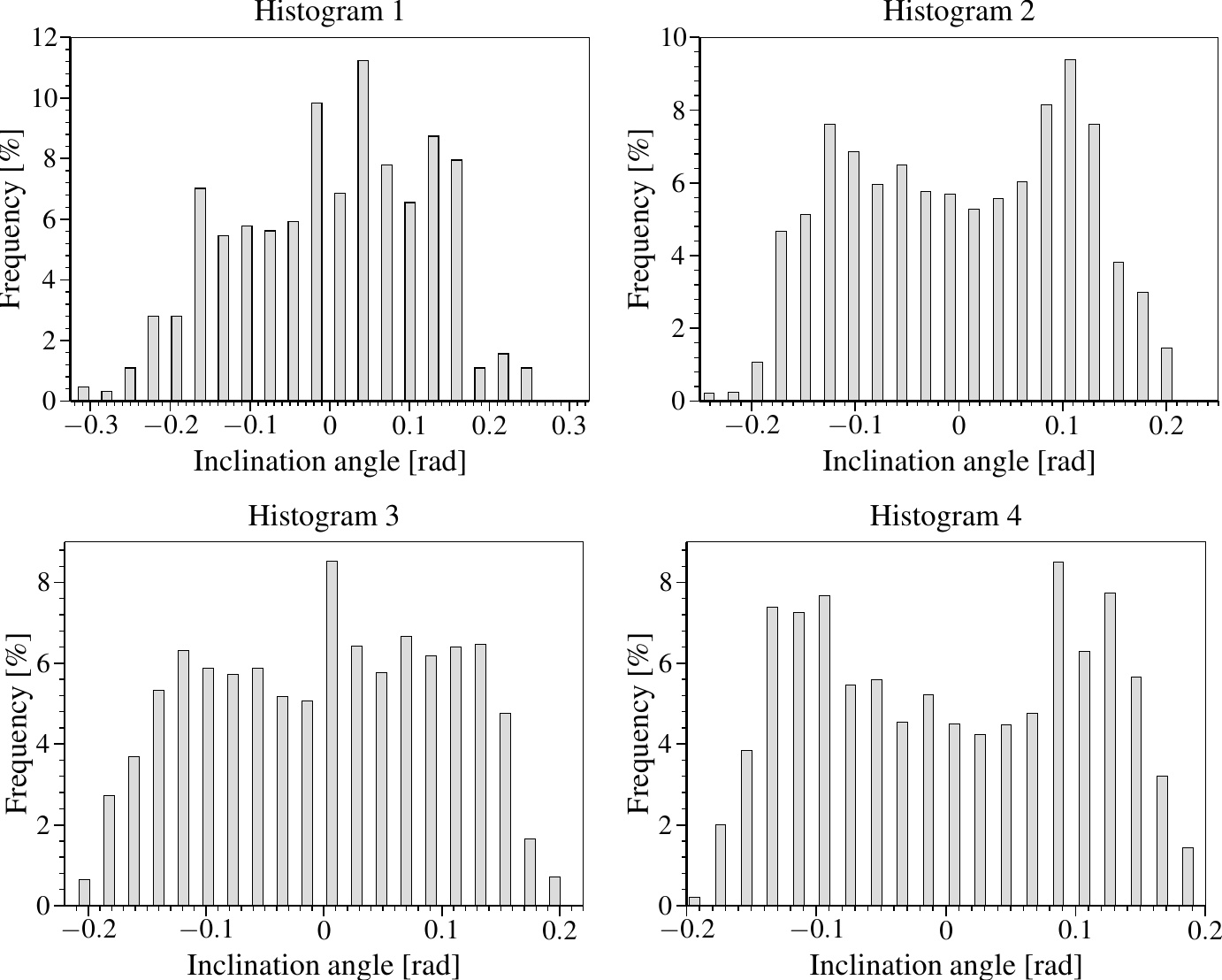}}
\caption{Real measured histograms of inclination
  angles~\cite{Vopicka:2004:PGV}.}
\label{fig5:hist}
\end{figure}

Having identified the optimal form of the Eshelby matrix, the
attention is now focused again on the general formulation and the
orientation averaging in particular, recall
\Eref{orient_average}. Unlike in \Sref{sec:mt-ideal}, however, the
joint probability density function is represented by real histograms
of the fiber tow orientation angles already introduced in
\Fref{F:EPUC}(f); see also \Fref{fig5:hist} for additional
examples. With such probabilistic characterization in hand, the warp
stiffness can be estimated as, cf.~\Eref{eq:warp},
\begin{equation}
\oavg{\mtrx{L}^{(1,{\rm warp})}} 
= 
\sum_{i=1}^{m} 
p_i \mtrx{L}^{(1)}( 0, \theta_i, 0 ),
\end{equation}
where $m$ denotes the number of sampling values and the discrete
angles $\theta_i$ and probabilities $p_i$ follow directly from the
image analysis data. The rest of the analysis exactly duplicates the
perfect unit cell case.

The complete data from the analyzed \CC~sample involve eleven such
histograms describing the waviness of the fiber tow in individual
plies. The final homogenized properties together with the elementary
statistical characterization are summarized in \Tref{T:MT-histograms}.

%%%%%%%%%%%%%%%%%%%%%%%%%%%%%%%%%%%%%%%%%%%%%%%%%%%%%%%%%%%%%%%%%%%%%%%%%%%%%%%%%%%%%%%%%%%%%%%%%%%%%%%%%%%%%%%%%%%%%%%%%%%%%%%%%%%%%%%%%%%%%%%%%%%%%%
\begin{table}[ht]
\centering
\caption{Homogenized effective stiffnesses determined by the
  \MT~scheme for \new{\CC~system and} histograms measured
  in~\cite{Vopicka:2004:PGV}.}
\label{T:MT-histograms}
\begin{tabular}{lrrrrrr}
\br
Histogram & 
$L^{\rm \MT}_{11}$ & $L^{\rm \MT}_{12}$ & $L^{\rm \MT}_{13}$ & 
$L^{\rm \MT}_{33}$ & $L^{\rm \MT}_{44}$ & $L^{\rm \MT}_{66}$ \\ 
& [GPa] & [GPa] & [GPa] & [GPa] & [GPa] & [GPa] \\
\mr
1  & 86.94 & 16.71 & 15.30 & 50.28 & 21.63 & 42.81 \\
2  & 88.19 & 17.16 & 15.64 & 51.15 & 22.23 & 44.27 \\
3  & 88.34 & 17.14 & 15.67 & 51.20 & 22.24 & 44.49 \\
4  & 86.73 & 16.64 & 15.24 & 50.11 & 21.49 & 42.58 \\
5  & 86.25 & 16.47 & 15.11 & 49.45 & 21.27 & 42.02 \\
6  & 86.57 & 16.58 & 15.20 & 49.99 & 21.63 & 42.39 \\
7  & 88.52 & 17.28 & 15.72 & 51.35 & 22.35 & 44.69 \\
8  & 90.26 & 17.90 & 16.20 & 52.59 & 23.23 & 46.71 \\
9  & 86.11 & 16.42 & 15.07 & 49.69 & 21.21 & 41.83 \\
10 & 86.82 & 16.67 & 15.26 & 50.17 & 21.54 & 42.68 \\
11 & 87.96 & 17.07 & 15.58 & 51.00 & 22.15 & 43.98 \\
\mr
Average   & 87.52 & 16.91 & 15.45 & 50.63 & 21.91 & 43.49 \\
Standard deviation  & 1.26 & 0.44 & 0.34 & 0.91 & 0.60 & 1.48 \\
\br
\end{tabular}
\end{table}

It becomes clear by associating the results in \Tref{T:MT-histograms}
with the corresponding micrographs that those segments which have more
fibers oriented near the direction $\theta = 0$ provide a stiffer
response than the others. As an illustration, consider e.g. histograms
No.~$2$ and $3$ in \Fref{fig5:hist} and corresponding stiffnesses in
\Tref{T:MT-histograms}. Nevertheless this difference is, due to a
rather narrow range of the inclination angles, not too important,
especially if concerning the approximate character of the \MT~method.

\section{Conclusions}\label{sec:con}
%%%%%%%%%%%%%%%%%%%%
%
In the present work, an efficient numerical method for the
homogenization of plain weave composites with both ideal and imperfect
tow paths based on the Mori-Tanaka method has been proposed. The
adopted strategy builds on the matching of results of the detailed FEM 
analysis with the micromechanical model. The most
pertinent conclusions can be stated as follows:

\begin{itemize}

\item[i)] The simplified method is able to deliver the homogenized
  parameters with values comparable with the detailed finite element
  simulations. The resulting method compares well with independent
  numerical approaches \replace{as well as}{and} available
  experimental data.

\item[ii)] The accuracy of the method depends on the shape of an
  equivalent ellipsoid, which represents both geometrical and
  mechanical effects, such as inter-tow interactions. The parameters
  of the inclusion follow from a well-defined global optimization
  problem and a FEM-generated training set. Moreover, the optimal
  shape is robust with respect to moderate geometry perturbations.

\item[iii)] The method allows us to directly assess the effects of tow
  waviness quantified by histograms of inclination angles, including
  the statistical characterization of the homogenized stiffnesses.

\end{itemize}

The future extension of the method will include the treatment of the
intrinsic porosity of the \CC~composite evident from
\Fref{F:EPUC}(a). In the framework of \new{multi-phase} Mori-Tanaka
approaches, the porosity can be modeled as an additional phase
characterized in the simplest case by volume fractions or by
three-dimensional computer tomography data, see
\new{\cite{Piat:2006:MMCVI_1,Piat:2006:MMCVI_2} for related
  studies}. Such work is in progress and will be reported separately.

\section*{Acknowledgments}
%%%%%%%%%%%%%%%%%%%%%%%%%%%%%%%%%%%%%%%%%%%%%%%%%%%%%%%%%%%%%%%%%%%%%%%%%%%%%%%%%%%%%%%%%%%%%%%%%%%%%%%%%%%%%%%%%%%%%%%%%%%%%
Authors would like to thank Jan Vorel for helpful discussions on the
subject. The financial support provided by the GA\v{C}R grant
No.~106/07/1244 and partially also by the research project
CEZ~MSM~6840770003 is gratefully acknowledged.

%%%%%%%%%%%%%%%%%%%%%%%%%%%%%%%%%%%%%%%%%%%%%%%%%%%%%%%%%%%%%%%%%%%%%%%%%%%%%%%%%%%%%%%%%%%%%%%%%%%%%%%%%%%%%%%%%%%%%%%%%%%%%

%\bibliographystyle{jphysicsB} 
%\bibliography{liter}

\appendix

\section{Global optimization algorithm}\label{sec:app}
%%%%%%%%%%%%%%%%%%%%%%%%%%%%%%%%%%%%%%%

\begin{figure}[ht]
\centerline{\figName{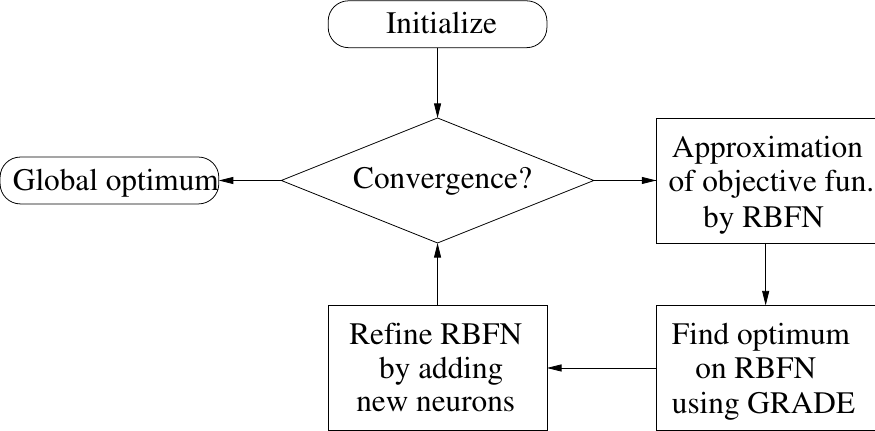}}
\caption{Flow chart of the applied algorithm.}
\label{flow_chart}
\end{figure}

The computation scheme is based on the genetic algorithm
GRADE~\cite{Ibrahimbegovic:2004:ODOC} evaluating the Radial Basis
Function Network (RBFN) approximation of the objective function,
see~\cite{Skocek:2005:Topping} for more detailed description. The
algorithm is briefly described in the flow chart depicted in
\Fref{flow_chart}.

In particular, instead of directly evaluating the objective function
$F( \sa_2, \sa_3 )$ defined by~\Eref{eq:obj_func}, the GA evaluates
its RBFN approximation. When the optimum of the approximation is
found, the RBFN is enriched with new neurons according to steps
described in~\cite{Skocek:2005:Topping} and the approximation is
refined. At this time the real objective function is evaluated at
several points. This cycle is repeated until the two consecutive
solutions differ by less than a certain specified value, set to
$10^{-2}$. Moreover, due to the intrinsic randomness of the algorithm,
all reported optimization results correspond to the optimum of five
independent executions.

\end{document}

%% file: format.ltx
\newcommand\de[1]{\,{\mathrm d}#1}

\newcommand{\bmath}[1]{\mbox{\boldmath$#1$}}

\newcommand{\mtrx}[1]{\bmath{#1}}

\newcommand{\avgs}[1]{\left\langle #1 \right\rangle} % Spatial average

\newcommand{\trn}{^{\sf T}}
\newcommand{\strain}{\varepsilon}
\newcommand{\stress}{\sigma}